\documentclass[conference,a4paper]{IEEEtran}
\usepackage{amsfonts,booktabs}
\usepackage{amsmath}
\usepackage{graphicx}
\usepackage{subfigure}
\usepackage{times}
\usepackage{pgfplots}
\usepackage{pgfplotstable}
\usepackage[subnum]{cases}
\usepackage{filecontents}
\usepackage{multirow}
\usepackage{soul}
\usepackage{color}
\usepackage{pdfsync}

\newtheorem{sideremark}{Remark}

\newtheorem{definition}         {Definition}[section]

\newtheorem{theorem}{Theorem}

\newcommand{\beq}{\begin{equation}}
\newcommand{\eeq}{\end{equation}}
\newcommand{\beqa}{\begin{eqnarray}}
\newcommand{\eeqa}{\end{eqnarray}}

\newcommand{\mathbi}[1]{\ensuremath \textbf{\em #1}}
\newcommand{\paren}[1]{\left(#1\right)}
\newcommand{\sqparen}[1]{\left[#1\right]}
\newcommand{\brparen}[1]{\left\{#1\right\}}
\newcommand{\field}[1]{\ensuremath{\mathbb{#1}}}
\newcommand{\R}{\ensuremath{\field{R}}} 
\newcommand{\Rp}{\ensuremath{\R_+}} 
\newcommand{\I}[1]{\ensuremath{\mathsf{1}_{\left\{#1\right\}}}} 
\newcommand{\Inb}[1]{\ensuremath{\mathsf{1}_{#1}}} 
\newcommand{\ra}{\ensuremath{\rightarrow}} 
\newcommand{\PR}[1]{\ensuremath{\mathsf{Pr}\left\{#1\right\}}} 
\newcommand{\PRP}[1]{\ensuremath{\mathsf{Pr}\left(#1\right)}} 
\newcommand{\EW}{\ensuremath{\mathsf{E}}} 
\newcommand{\ES}[1]{\ensuremath{\mathsf{E}\left[#1 \right]}} 
\newcommand{\e}[1]{\ensuremath{{\rm e}^{#1}}} 

\newcommand{\LO}[1]{\ensuremath{o\paren{#1}}}

\newcommand{\logp}[1]{\ensuremath{\log\paren{#1}}}

\newcommand{\CRTPIL}{\ensuremath{\frac{h_i}{\lambda_N+\mu_Ng_i}}}

\newcommand{\CRDTPIL}{\ensuremath{F_{\lambda_N,\mu_N}\paren{x}}}
\newcommand{\ICRDTPIL}[1]{\ensuremath{F^{-1}_{\lambda_N,\mu_N}\paren{1-#1}}}
\newcommand{\PTPIL}{\ensuremath{P\paren{h_i,g_i}}}
\newcommand{\Xs}[2]{\ensuremath{X_N^\star\paren{#1,#2}}}
\newcommand{\Xd}[2]{\ensuremath{X_N^\diamond\paren{#1,#2}}}
\newcommand{\RTPIL}[1]{\ensuremath{R\paren{#1N}}}

\newcommand{\Sch}[2]{\ensuremath{S\paren{#1,#2}}}

\newcommand{\pfgsetting}{\pgfplotsset{
width= 8cm, 
every axis/.append style={line width=1.2pt},
label style={font=\bf\scriptsize}, 
ylabel style={yshift=-0.8em},
xlabel={Number of Secondary Users},
ylabel={Throughput [nats/s/Hz]},
title style={font=\bf\scriptsize}, 
tick label style={font=\scriptsize,/pgf/number format/1000 sep={} },
tick style={ line width=1.5pt},
legend style={font=\bf\tiny,cells={anchor=west}},
every mark/.append style={solid}
}}
\newcommand{\axissetting}{ xmin=0,xmax=1000,xtick={0 ,100,200,300,400,500,600,700,800,900,1000},grid=major, tick style={color=black, major tick length={0.10 cm}}, grid style={line width= 0.75pt, densely dotted, color= black}}

\begin{document}

\sloppy

\title{Distributed Cognitive Multiple Access Networks: Power Control, Scheduling and Multiuser Diversity}

\author{
  \IEEEauthorblockN{Ehsan Nekouei}
  \IEEEauthorblockA{Dept. of Electrical \& Electronic Eng.\\
    The University of Melbourne\\
    Victoria, Australia\\
    Email: e.nekouei@student.unimelb.edu.au} 
  \and
  \IEEEauthorblockN{Hazer Inaltekin}
  \IEEEauthorblockA{Dept. of Electrical \& Electronics Eng.\\
    Antalya International University\\
    Antalya, Turkey\\
    Email: hazeri@antalya.edu.tr}
  \and
  \IEEEauthorblockN{Subhrakanti Dey}
  \IEEEauthorblockA{Dept. of Electrical \& Electronic Eng.\\
    The University of Melbourne\\
    Victoria, Australia\\
    Email: sdey@unimelb.edu.au}
}



\maketitle

\begin{abstract}
This paper studies optimal distributed power allocation and scheduling policies (DPASPs) for distributed total power and interference limited (DTPIL) cognitive multiple access networks in which secondary users (SU) independently perform power allocation and scheduling tasks using their local knowledge of secondary transmitter secondary base-station (STSB) and secondary transmitter primary base-station (STPB) channel gains.  In such networks, transmission powers of SUs are limited by an average total transmission power constraint and by a constraint on the average interference power that SUs cause to the primary base-station.  We first establish the joint optimality of water-filling power allocation and threshold-based scheduling policies for DTPIL networks. We then show that the secondary network throughput under the optimal DPASP scales according to $\frac{1}{\e{}n_h}\log\logp{N}$, where $n_h$ is a parameter obtained from the distribution of STSB channel power gains and $N$ is the total number of SUs.  From a practical point of view, our results signify the fact that distributed cognitive multiple access networks are capable of harvesting multiuser diversity gains without employing centralized schedulers and feedback links as well as without disrupting primary's quality-of-service (QoS).
\end{abstract}

\section{Introduction}
Traditional command-and-control approach towards spectrum allocation has already been a widely recognized bottleneck for the successful management of exponentially growing data-hungry wireless communication services \cite{Haykin05}-\cite{Akyildiz06}. Exploiting spatially and temporally underutilized frequency bands, cognitive radio technology is a promising solution for improving spectral efficiency of next generation wireless networks. To solve the spectrum scarcity problem, cognitive radio technology utilizes a suite of efficient communication techniques and approaches that enable secondary users (SUs) (or, alternatively called cognitive users) to access the same frequency band of primary users (PUs) while PUs' quality-of-service (QoS) is protected to be above some certain level. To this  end, interference management task is of prime importance for the cognitive radio network design.

Due to dependence of the interference management task on the knowledge of channel gains between secondary transmitters and primary receivers, channel state information (CSI) plays a critical role in the design and successful implementation of cognitive radio protocols. On the other hand, availability of \emph{centralized} CSI at the secondary network relies on the existence of feedback links for conveying CSI.  It also highly depends on the inherent physical characteristics of wireless channels such as capacity and channel coherence time. Hence, assumptions on the availability of centralized CSI at the secondary network may not be realistic for practical cognitive radio networks. However, optimal resource allocation and capacity limits of cognitive radio networks have been mainly studied under the full or partial CSI assumption in the literature, {\em e.g.,} see \cite{RZhang09}-\cite{cogmud_alitajer10}.   

In \cite{cogmud_twban09}, Ben et al. established the logarithmic and double-logarithmic capacity scaling laws for cognitive multiple access networks under peak transmission power and peak interference power constraints for Rayleigh fading channels.  Similar results were obtained for cognitive multiple access, cognitive broadcast and cognitive parallel access channels in \cite{cogmid_zhang10}. In \cite{NID12}, the authors established the logarithmic and double logarithmic throughput scaling laws for interference-limited and total-power-and-interference-limited cognitive multiple access networks when transmission powers of SUs are optimally allocated. In \cite{NIDSubmitted}, the throughput scaling behavior of cognitive multiple access networks was investigated when interference channel gains are partially available at the secondary network and the distributions of channel gains belong to a fairly large class of distribution functions called class-$\mathcal{C}$ distribution functions. 

In \cite{cogmudmsd_10}, the authors obtained capacity expressions for a cognitive broadcast network sharing multiple orthogonal frequency bands with a primary network under peak interference power constraints in each band. In \cite{cogmud_alitajer10}, a secondary network with  $N$ secondary transmitter-receiver pairs sharing $M$ frequency bands with a primary network was considered. The authors established a double-logarithmic scaling law for the secondary network capacity under the optimum matching of $M$ SUs with $M$ primary network frequency bands. They also established the double-logarithmic scaling law for the secondary network capacity under a contention-free distributed scheduling algorithm without providing any protection for primary's transmission for Rayleigh fading channels.

Different from previous works, this paper studies the optimal distributed power allocation and scheduling policies (DPASPs) for distributed total power and interference limited (DTPIL) cognitive multiple access networks. In DTPIL networks, transmission powers of SUs are limited by a constraint on the average total transmission power of the secondary network and a constraint on the average interference power of the secondary network at the primary base-station (PBS). In DTPIL networks, each SU obtains the knowledge of its secondary transmitter secondary base-station (STSB) and secondary transmitter primary base-station (STPB) channels gains using pilot signals transmitted intermittently by the SBS and PBS.  After acquiring knowledge of its own STSB and STPB channel gains, each SU employs this information to perform the scheduling and power allocation tasks without any feedback from the secondary base-station (SBS). 

Assuming a collision channel model for secondary transmissions, we first show that the jointly optimal power allocation and scheduling policy is in the form of water-filling power allocation and threshold-based scheduling.  Then, under the optimal DPASP, we study the throughput scaling behavior of DTPIL networks when distributions of STSB and STPB channel power gains belong to class-$\mathcal{C}$ distribution functions \cite{NIDSubmitted}.  It is shown that the secondary network throughput scales according to $\frac{1}{\e{}n_h}\log\logp{N}$, where $n_h$ is a parameter obtained from the distribution of STSB channel power gains. $n_h$ is equal to $\frac{c}{2}$\footnote{$c$ is the Weibull distribution parameter \cite{Simon-Alouini05}.} for Weibull distributed STSB channel gains, and equal to$1$ for Rayleigh, Rician and Nakagami-$m$ distributed STSB channel gains . These results indicate that the cognitive radio networks are capable of performing interference management, power allocation and scheduling tasks in a distributed fashion as well as exploiting multiuser diversity gain.
 
The rest of the paper is organized as follows.  In the next section, we describe the system model and our modeling assumptions. In Section \ref{Sec: OP}, we derive the structure of optimal DPASPs.  Our capacity scaling results are presented in Section \ref{Sec: R&D} along with some numerical figures.  Section \ref{sec: conclusion} concludes the paper.

\section{System Model}\label{Sec: System Model}
\begin{figure}[!t]
\centering{\includegraphics[scale=0.38]{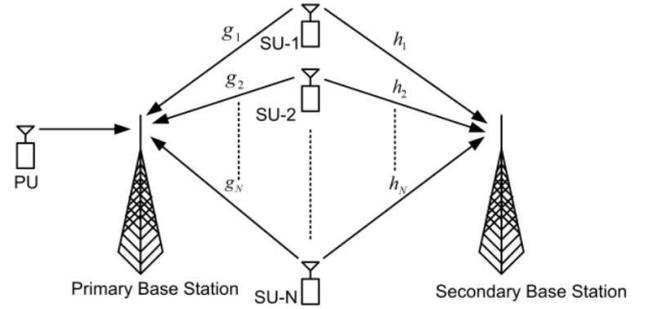}}
\caption{A cognitive multiple access network with $N$ SUs.} \label{F1}
\end{figure}

We consider a cognitive multiple access network in which $N$ SUs communicate with an SBS and simultaneously cause interference to a PBS as depicted in Fig. \ref{F1}. Let $h_i$ and $g_i$ represent the $i$th STSB and STPB channel power gains, respectively. We consider the classical ergodic block fading model \cite{Tse} to model the statistical variations of all STSB and STPB channel gains. $\left\{h_i\right\}_{i=1}^N$ and $\left\{g_i\right\}_{i=1}^N$ are assumed to be collections of i.i.d. random variables distributed according to distribution functions $F_h\paren{x}$ and $F_g\paren{x}$, respectively. The random vectors ${\mathbi h}=\sqparen{h_1,h_2,\ldots,h_N}^\top$ and ${\mathbi g}= \sqparen{g_1,g_2,\ldots,g_N}^\top$ are assumed to be independent from each other.  We assume that each SU has access to its STSB and STPB channel gains by means of pilot training signals periodically transmitted by the SBS and PBS \cite{Berry06}.  
\begin{definition}\label{Def1}
We say that the cumulative distribution function (CDF) of a random variable $X$, denoted by $F\paren{x}$, belongs to the class $\cal C$-distributions if it satisfies the following properties:
\begin{itemize}
\item $F\paren{x}$ is continuous.
\item $F(x)$ has positive support, {\em i.e.,} $F(x)=0$ for $x \leq 0$.
\item $F(x)$ is strictly increasing, {\em i.e.,} $F(x_1)<F(x_2)$ for $0<x_1<x_2$.
\item The tail of $F(x)$ decays to zero \emph{double exponentially}, {\em i.e.,} there exist constants $\alpha>0$, $\beta>0$, $n>0$, $l \in \R$ and a slowly varying function $H(x)$ satisfying $H(x) = \LO{x^{n}}$ such that   
  \begin{eqnarray}\label{Eq: tail-condition-1}
 \lim_{x\ra\infty}\frac{1-F(x)}{\alpha x^{l}\e{\paren{-\beta x^{n}+H(x)}}}=1.
 \end{eqnarray}
\item $F(x)$ varies \emph{regularly} around the origin, {\em i.e.,} there exist constants $\eta>0$ and $\gamma>0$ such that
  \begin{eqnarray}\label{Eq: tail-condition-2}
 \lim_{x\downarrow0}\frac{F(x)}{\eta x^{\gamma}}=1.
 \end{eqnarray}       
\end{itemize}
  \end{definition}

In this paper, we assume that the CDFs of all fading power gains belong to the class $\cal C$-distributions. In Table \ref{Table: Fading Parameters}, we illustrate the parameters characterizing the behavior of the distribution of fading power gains around zero and infinity for the commonly used fading models in the literature. To avoid any confusion, these parameters are represented by subscript $h$ for STSB channel gains and with subscript $g$ for STPB channel gains in the sequel.

\begin{table*}[!t]
\begin{minipage}{\textwidth}
\renewcommand{\arraystretch}{1.3}
\caption{Common fading channel models and their parameters}
\centering
\begin{tabular}{cccccccc}
\toprule
\multicolumn{1}{c}{\multirow{2}{*}{ Channel Model}}  & \multicolumn{7}{c}{ Parameters }\\
 \cmidrule(r){2-8}
&  $\alpha$ & $l$& $\beta$ & $n$ & $H(x)$ & $\eta$ & $\gamma$  \\
\midrule
Rayleigh & 1&  $0$ & 1 & 1 & 0 & 1&1 \\
\midrule
Rician & $\frac{1}{2\sqrt{\pi}\e{K_f}\sqrt[4]{K_f\paren{K_f+1}}}$ & $-\frac{1}{4}$ & $K_f+1$ &1 & $2\sqrt{K_f\paren{K_f+1}x}$& $\frac{K_f+1}{\e{K_f}}$ & 1\\
\midrule
Nakagami-$m$ & $\frac{m^{m-1}}{\Gamma(m)}$ & $m-1$ & $m$ & 1 & 0& $\frac{m^{m-1}}{\Gamma(m)}$& $m$ \\
\midrule
Weibull & 1& 0 & $\Gamma^\frac{c}{2}\paren{1+\frac{2}{c}}$ & $\frac{c}{2}$ & $0$ & $\Gamma^\frac{c}{2}\paren{1+\frac{2}{c}}$ & $\frac{c}{2}$ \\\bottomrule
\label{Table: Fading Parameters}
\end{tabular}
\end{minipage}
\end{table*}

In DTPIL networks, each SU exploits the knowledge of its STSB and STPB channel gains to perform scheduling and power allocation tasks \emph{locally} and \emph{independently} from other SUs due to lack of a centralized scheduler and feedback links.  In this setting, we define the power allocation policy $P\paren{\cdot,\cdot}$ as a mapping from $\R_+^2$ to $\R_+$, where $P\paren{h_i,g_i}$ represents the transmission power of the $i$th SU at the joint channel state $\paren{h_i, g_i}$.  We also define the scheduling function $S\paren{\cdot,\cdot}$ as a mapping from $\R_+^2$ to $\left\{0,1\right\}$, where $S\paren{h_i,g_i}=1$ means that the $i$th SU transmits with power $P(h_i, g_i)$ at the joint channel state $\paren{h_i, g_i}$, and remains silent otherwise. The scheduling function is designed such that the scheduling probability for transmission is equal to $p$, $p \in (0, 1)$, for all SUs, \emph{i.e.,} $\PR{S\paren{h_i,g_i}=1} = p$ for $i\in\left\{1,\cdots,N\right\}$.  We also assume that the background noise power is normalized to $1$, and given a joint channel state $\paren{h_i, g_i}$,  the $i$th SU transmits at rate $\logp{1+h_i\PTPIL}\I{\Sch{h_i}{g_i}=1}$ [nats/s/Hz].  Hence, if two or more SUs transmit concurrently, the SBS cannot decode any data stream, declares a collision and the resulting throughput becomes equal to zero.  Here, $p$ parameter can be considered as a design degree-of-freedom helping us to keep collisions below some certain level. Under these modeling assumptions, the average total transmission power and the average interference power at the PBS can be expressed as in \eqref{Eq: Power} and \eqref{Eq: Interference}, respectively,
\begin{figure*}[t]
\begin{eqnarray}\label{Eq: Power}
\ES{\sum_{i=1}^NP\paren{h_i,g_i}\I{\Sch{h_i}{g_i}=1}}=N\ES{P\paren{h,g}\I{\Sch{h}{g}=1}}
\end{eqnarray}
\hrule
\end{figure*}
\begin{figure*}[t]
\begin{eqnarray}\label{Eq: Interference}
\ES{\sum_{i=1}^Ng_iP\paren{h_i,g_i}\I{\Sch{h_i}{g_i}=1}}=N\ES{gP\paren{h,g}\I{\Sch{h}{g}=1}}
\end{eqnarray}
\hrule
\end{figure*}
where $h$ and $g$ are two independent generic random variables distributed according to $F_h\paren{x}$ and $F_g\paren{x}$, respectively. 

\section{Optimal power allocation and scheduling policies}\label{Sec: OP}
In this section, we  will derive jointly optimal power allocation and scheduling policies maximizing the transmission rates of SUs subject to average total power, average interference power and scheduling probability for transmission constraints. Formally speaking, we look for the solutions of the following functional optimization problem:
\begin{eqnarray}\label{OP}
\begin{array}{ll}
\underset{P\paren{h,g},S\paren{h,g}}{\mbox{maximize}} &\EW_{h, g} \sqparen{ \log\paren{ 1+  hP\paren{h,g} }\I{S\paren{h,g}=1} } \\
\mbox{subject to} & \EW_{h,g}\sqparen{P\paren{h,g}\I{S\paren{h,g}=1} } \leq \frac{P_{\rm ave}}{N} \\
			    & \EW_{h,g} \sqparen{gP\paren{h,g}\I{S\paren{h,g}=1}} \leq \frac{Q_{\rm ave}}{N}\\
			    &\PR{S\paren{h,g} = 1} = p
\end{array}, \label{macprob}
\end{eqnarray}
where $P_{\rm ave}$ and $Q_{\rm ave}$ are total transmission and interference power constraints, respectively.  
In the next theorem, we establish the structure of jointly optimal power allocation and scheduling policies solving \eqref{macprob}.  We note that the optimization problem \eqref{macprob} is not convex due to the scheduling probability for transmission constraint. However, in the proof of the next theorem, we solve \eqref{macprob} by approximating it from above through a convex formulation, and showing that the upper bound can be achieved by a feasible point of \eqref{macprob}.    
\begin{theorem}\label{Theo: Opt}
Let the $P^\star\paren{h,g}$ and $S^\star\paren{h,g}$ be a solution of \eqref{macprob}.  Then, 
 \begin{eqnarray} 
P^\star\paren{h,g} &=& \paren{\frac{1}{\lambda_N+\mu_Ng}-\frac{1}{h}}^+\nonumber
\end{eqnarray}
and
\begin{eqnarray}
S^\star\paren{h,g} &=& \I{\frac{h}{\lambda_N+\mu_N g}>\ICRDTPIL{p}}, \label{Eq: TPIL-powalloc}
\end{eqnarray}
where $F^{-1}_{\lambda_N,\mu_N}\paren{x}$ is the functional inverse of the CDF of $\frac{h}{\lambda_N+\mu_N g}$, \emph{i.e.,} $\CRDTPIL$, $\lambda_N$ and $\mu_N$ are Lagrange multipliers associated with average total transmission and interference power constraints in \eqref{OP-Auxiliary-1}, respectively.
\end{theorem}
\begin{IEEEproof}
Please see the Appendix.
\end{IEEEproof}

Theorem \ref{Theo: Opt} implies that the $i$th SU schedules its transmission using a water-filling power allocation policy if its joint power and interference channel state, \emph{i.e.,} $\CRTPIL$, is above the threshold value of $\ICRDTPIL{p}$. Let $\Xs{\lambda_N}{\mu_N}$ and $\Xd{\lambda_N}{\mu_N}$ be the largest and the second largest elements among the collection of i.i.d random variables $\brparen{X_i\paren{\lambda_N,\mu_N}}_{i=1}^N$, respectively, where $X_i\paren{\lambda_N,\mu_N}=\CRTPIL$. Then, the sum-rate in DTPIL networks can be expressed as 
\begin{eqnarray}
\RTPIL{}=\ES{\logp{\Xs{\lambda_N}{\mu_N}}\Inb{A_N}},
\end{eqnarray}
 where 
 \begin{eqnarray}
 \lefteqn{A_N=\left\{\Xs{\lambda_N}{\mu_N}>\max\paren{\ICRDTPIL{p},1},\right.}\hspace{9cm}\nonumber\\
 \lefteqn{\left.\Xd{\lambda_N}{\mu_N}\leq \max\paren{\ICRDTPIL{p}, 1}\right\}.}\hspace{6.7cm}\nonumber
 \end{eqnarray}
 
\section{Capacity Scaling Result}\label{Sec: R&D}
Now, we study the capacity scaling behavior of DTPIL networks under optimal DPASPs along with some discussions and numerical studies. For simplicity, we set $p = \frac{1}{N}$ in the remainder of the paper.  Next theorem provides the throughput scaling law for DTPIL networks.
\begin{theorem}\label{Theo: DTPIL}
Let $\RTPIL{}$ be the secondary network sum-rate in DTPIL networks under optimal DPASPs. Then, 
\begin{eqnarray}
\lim_{N\ra\infty}\frac{\RTPIL{}}{\log\logp{N}}=\frac{1}{\e{} n_h}.
\end{eqnarray}
\end{theorem}
\begin{IEEEproof}
Please see \cite{NIDTBSubmitted}.
\end{IEEEproof}

Theorem \ref{Theo: DTPIL} establishes the double logarithmic throughput scaling law for DTPIL networks under optimal DPASPs when transmission probabilities of SUs are all equal to $\frac{1}{N}$.  Theorem \ref{Theo: DTPIL} also reveals that the secondary network throughput in DTPIL networks is affected by a pre-log factor of $\frac{1}{\e{} n_h}$. $n_h$ is equal to $\frac{c}{2}$ for Weibull distributed STSB channel gains and equal to $1$ for Rayleigh, Rician and Nakagami-$m$ distributed STSB channel gains. The result of Theorem \ref{Theo: DTPIL} can be intuitively explained as follows.  Note that $\PRP{A_N}$ represents the fraction of time that only the SU with the maximum of $\CRTPIL$ transmits.  It can be easily verified that $\PRP{A_N}$ converges to $\frac{1}{\e{}}$ as $N$ tends to infinity. Hence, as the number of SUs becomes large, the fraction of time that just the best SU transmits is approximately equal to $\frac{1}{\e{}}$.  It is also shown in \cite{NIDTBSubmitted} that $\logp{\Xs{\lambda_N}{\mu_N}}$ scales according to $\frac{1}{n_h}\log\logp{N}$. These observations suggest that the secondary network throughput should scale according to $\frac{1}{\e{}n_h}\log\logp{N}$ as $N$ becomes large, which is indeed the case.  
 
The secondary network sum-rate can also be written as 
\begin{eqnarray}\label{Eq: Rate-TPIL}
\lefteqn{R\paren{N}=\logp{\frac{1}{\lambda_N}}\ES{\Inb{A_N}}+}\hspace{8cm}\nonumber\\
\lefteqn{\ES{\logp{\Xs{1}{\frac{\mu_N}{\lambda_N}}}\Inb{A_N}}}\hspace{4.5cm}
\end{eqnarray}

It can be shown that the Lagrange multiplier associated with the average total power constraint, $\lambda_N$, converges to $\frac{1}{P_{\rm ave}}$ as $N$ becomes large. Thus, the first term in \eqref{Eq: Rate-TPIL} converges to $\frac{1}{\e{}}\logp{P_{\rm ave}}$ as $N$ becomes large, which indicates the logarithmic effect of the total power constraint on the secondary network throughput in DTPIL networks. 
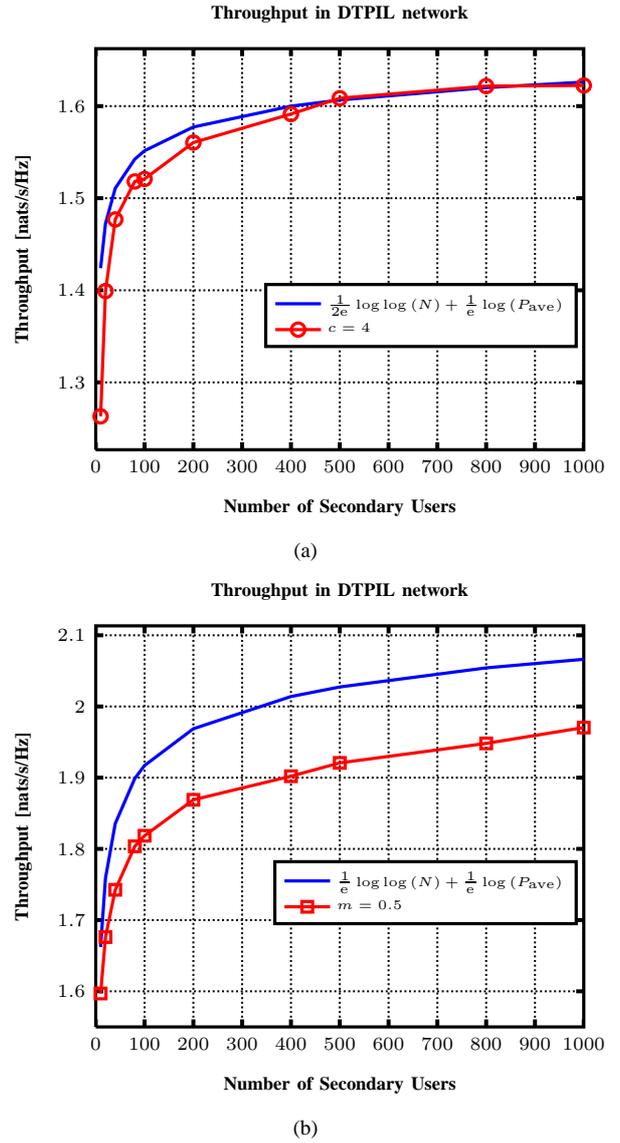
\begin{figure}[t]
\centering
\subfigure[]
{
\begin{tikzpicture}
\begin{axis}[title={Throughput in DTPIL network},legend style={yshift=-3cm, xshift=0.05cm},\axissetting]
                                      
\pfgsetting
\addplot+[color=blue,mark=none] table [x=N,y=0.5loglog]{WeiRTPIL.dat};\addlegendentry{$\frac{1}{2\e{}}\log\logp{N}+\frac{1}{\e{}}\logp{P_{\rm ave}}$};
\addplot+[mark=o, mark size=2.5pt] table [x=N,y=c4]{WeiRTPIL.dat};\addlegendentry{$c=4$};
\end{axis}
\end{tikzpicture}
\label{WeiRTPIL2}}
\subfigure[]
{
\begin{tikzpicture}
\begin{axis}[title={Throughput in DTPIL network},legend style={yshift=-3cm, xshift=0.05cm},\axissetting]
                                      
\pfgsetting
\addplot+[color=blue,mark=none] table [x=N,y=loglog]{RNaTPIL.dat};\addlegendentry{$\frac{1}{\e{}}\log\logp{N}+\frac{1}{\e{}}\logp{P_{\rm ave}}$};
\addplot+[color=red,mark=square] table [x=N,y=m0.5]{RNaTPIL.dat};\addlegendentry{$m=0.5$};
\end{axis}
\end{tikzpicture}
\label{RNaTPIL}}
\caption{ Secondary network throughput in DTPIL networks with increasing numbers of SUs. $P_{\rm ave}$ and $Q_{\rm ave}$ are set to 15dB and 0dB, respectively.}
\label{FTPIL}
\end{figure}

Figure \ref{FTPIL} demonstrates the secondary network throughput scaling behavior with increasing numbers of SUs in DTPIL networks for different communication environments. In this figure, $P_{\rm ave}$ and $Q_{\rm ave}$ are set to 15dB and 0dB, respectively. Similar qualitative behavior continues to hold for other values of $P_{\rm ave}$ and $Q_{\rm ave}$. In Fig. \ref{WeiRTPIL2}, STSB channel gains are Weibull distributed with $c=4$ and STPB channel gains are Rayleigh distributed.  As Fig. \ref{WeiRTPIL2} shows, the secondary network throughput scales according to $\frac{2}{\e{} \cdot c}\log\logp{N}$ with increasing numbers of SUs, \emph{i.e.,} $\frac{1}{2\e{}}\log\logp{N}$ for $c=4$, as predicted by Theorem \ref{Theo: DTPIL}. In Fig. \ref{RNaTPIL}, STSB channel gains are Rayleigh distributed and STPB channel gains are Nakagami-$m$ distributed with Nakagami parameter $m=0.5$.  As Fig. \ref{RNaTPIL} shows, the secondary network throughput scales according to $\frac{1}{\e{}}\log\logp{N}$ with increasing numbers of SUs, which is also in accordance with Theorem \ref{Theo: DTPIL}.  Also, closeness of simulated data rates to the curve $\frac{1}{2\e{}}\log\logp{N} + \frac{1}{\e{}}\logp{P_{\rm ave}}$ in Fig. \ref{WeiRTPIL2} and to the curve $\frac{1}{\e{}}\log\logp{N} + \frac{1}{\e{}}\logp{P_{\rm ave}}$ in Fig. \ref{RNaTPIL} further illustrates the logarithmic effect of $P_{\rm ave}$ on the secondary network throughput. 

\section{conclusion}\label{sec: conclusion}
In this paper, we have investigated the jointly optimal power allocation and scheduling policies as well as capacity scaling laws for DTPIL cognitive multiple access networks. In DTPIL networks, transmission powers of SUs are limited by an average total transmission power  constraint and by a constraint on the average interference power of SUs at the PBS. In this setting, SUs perform power allocation and scheduling tasks by using their local knowledge of STSB and STPB channel gains without any feedback from the SBS. Assuming a collision channel to model transmissions of SUs, it has been first shown that the water-filling power allocation and threshold-based scheduling policies are jointly optimal for DTPIL networks.  Then, it has been shown that the secondary network throughput under the optimal DPASPs scales according to $\frac{1}{\e{}n_h}\log\logp{N}$, where $n_h$ is a parameter obtained from the distribution of STSB channel power gains.
 
\appendix\label{App: A1}
Consider the following auxiliary optimization problem:
\begin{eqnarray}\label{OP-Auxiliary}
\begin{array}{ll}
\underset{P\paren{h,g},W\paren{h,g}}{\mbox{maximize}} & \EW_{h, g}\sqparen{W\paren{h,g} \log\paren{ 1+  hP\paren{h,g}}}\\
\mbox{subject to} & \EW_{h,g}\sqparen{W\paren{h,g}P\paren{h,g}} \leq \frac{P_{\rm ave}}{N} \\
			    & \EW_{h,g} \sqparen{W\paren{h,g}gP\paren{h,g}} \leq \frac{Q_{\rm ave}}{N}\\
			    &\EW_{h,g}\sqparen{W\paren{h,g}}= p \\
			    &0\leq W\paren{h,g}\leq 1 
\end{array}, 
\end{eqnarray}
 
 If a pair of power allocation $P\paren{h,g}$ and scheduling policy $S\paren{h,g}$ is feasible for \eqref{OP}, then $P\paren{h,g}$ and $W\paren{h,g}=S\paren{h,g}$ are also feasible for \eqref{OP-Auxiliary}.  Hence, the optimal values of \eqref{OP-Auxiliary} form an upper bound on the optimal values of \eqref{OP}.  
 Using the change of variable $Q\paren{h,g} = P\paren{h,g} W\paren{h,g}$, \eqref{OP-Auxiliary} can be transformed into the following convex optimization problem:
\begin{eqnarray}\label{OP-Auxiliary-1}
\begin{array}{ll}
\underset{Q\paren{h,g},W\paren{h,g}}{\mbox{maximize}} & \EW_{h, g}\sqparen{W\paren{h,g} \log\paren{ 1+  \frac{hQ\paren{h,g}}{W\paren{h,g}}}}\\
\mbox{subject to} & \EW_{h,g}\sqparen{Q\paren{h,g}} \leq \frac{P_{\rm ave}}{N} \\
			    & \EW_{h,g} \sqparen{gQ\paren{h,g}} \leq \frac{Q_{\rm ave}}{N}\\
			    &\EW_{h,g}\sqparen{W\paren{h,g}}= p \\
			    &0\leq W\paren{h,g}\leq 1
\end{array}, 
\end{eqnarray}

It can be shown that the objective function in \eqref{OP-Auxiliary-1} as a function of $Q$ and $W$ is concave on \Rp. The Lagrangian for \eqref{OP-Auxiliary-1} can be written as
\begin{eqnarray}
\lefteqn{L\paren{Q, W, \lambda,\mu,\eta}= W\paren{h,g} \log\paren{ 1+  \frac{hQ\paren{h,g}}{W\paren{h,g}}}}\hspace{9cm}\nonumber\\
\lefteqn{-\lambda Q\paren{h,g}- \mu gQ\paren{h,g} -\eta W\paren{h,g},}\hspace{5.8cm}\nonumber
\end{eqnarray}
where $\lambda\geq0$, $\mu\geq0$ and $\eta$ are Lagrange multipliers associated with the average transmit power, average interference power and scheduling probability for transmission constraints, respectively. Using generalized Karush-Kuhn-Tucker (KKT) conditions \cite{Yates2005}, we need to have 
\begin{eqnarray}
\lefteqn{\frac{\partial L\paren{Q, W^\star, \lambda,\mu,\eta}}{\partial Q\paren{h,g}} \Big|_{Q = Q^\star}} \hspace{9cm} \nonumber \\ 
\lefteqn{=\frac{h}{1+\frac{hQ^\star\paren{h,g}}{W^\star\paren{h,g}}}-\lambda -\mu \left\{
\begin{array}{cc}
 =0 &  Q^\star\paren{h,g}>0  \nonumber\\
 \leq 0 &  Q^\star\paren{h,g}=0 
\end{array},
\right.} \hspace{7.5cm}
\end{eqnarray}
which implies $P^\star\paren{h,g}=\paren{\frac{1}{\lambda+\mu g}-\frac{1}{h}}^+$. From KKT conditions, we also need to have
\begin{eqnarray}
\lefteqn{\frac{\partial L\paren{Q^\star, W, \lambda,\mu,\eta}}{\partial W\paren{h,g}} \Big|_{W = W^\star} =\logp{1+hP^\star\paren{h,g}}} \hspace{9cm} \nonumber \\
\lefteqn{-\lambda P^\star\paren{h,g} - \mu gP^\star\paren{h,g}-\eta
\left\{
\begin{array}{cc}
 =0 &  0<W^\star\paren{h,g}<1  \nonumber\\
 \leq 0 &  W^\star\paren{h,g}=0 \nonumber\\
  \geq 0 &  W^\star\paren{h,g}=1 
\end{array},
\right.}\hspace{9cm}\nonumber
\end{eqnarray}
For $\frac{\partial L\paren{Q^\star, W, \lambda,\mu,\eta}}{\partial W\paren{h,g}}=0$, we have $\logp{1+hP^\star\paren{h,g}}-\lambda P^\star\paren{h,g} -\mu gP^\star\paren{h,g}=\eta$, which happens with zero probability since fading channel gains have continuous distributions. Thus, $W^\star\paren{h,g}\in\left\{0,1\right\}$ with probability one. For $\frac{\partial L\paren{Q^\star, W, \lambda,\mu,\eta}}{\partial W\paren{h,g}}\geq0$, we have
\begin{eqnarray}\label{Eq: A}
\logp{1+hP^\star\paren{h,g}}-\lambda P^\star\paren{h,g} -\mu gP^\star\paren{h,g}-\eta\geq 0.
\end{eqnarray}
Substituting $P^\star\paren{h,g}$ in \eqref{Eq: A}, we have  
\begin{eqnarray}\label{Eq: B}
\paren{\logp{\frac{h}{\lambda+\mu g}}+\frac{\lambda+\mu g}{h}-1}\I{\frac{h}{\lambda+\mu g}\geq 1}\geq \eta.
\end{eqnarray}

Since $G\paren{x}=\logp{x}+\frac{1}{x}-1$ is monotonically increasing for $x \geq 1$, \eqref{Eq: B} implies that $W^\star$ can be chosen as $W^\star\paren{h,g}=1$ for $\frac{h}{\lambda+\mu g}\geq \ICRDTPIL{p}$, which completes  the proof.
\bibliographystyle{IEEEtran}

\end{document}